\documentclass[aps,onecolumn,preprintnumbers,nofootinbib,superscriptaddress]{revtex4}
\usepackage{amsmath}
\usepackage{graphicx}
\usepackage{amsfonts}
\usepackage{array}
\usepackage{amsthm}
\usepackage{bm}
\usepackage{palatino}
\usepackage{mathpazo}
\usepackage{supertabular}
\usepackage{subfig}
\usepackage[breaklinks]{hyperref}
\usepackage{color}
\usepackage[font=small,labelfont=bf,
justification=justified,
format=plain]{caption}
\usepackage{graphicx}
\usepackage{caption}
\usepackage[font=small,labelfont=bf,justification=justified]{caption}
\usepackage[labelfont=bf,labelsep=quad,justification=justified]{caption}
\usepackage{epstopdf}

\newcommand{\be}{\begin{equation}}
\newcommand{\ee}{\end{equation}}
\newcommand{\ba}{\begin{eqnarray}}
\newcommand{\ea}{\end{eqnarray}}
\newcommand{\bal}{\begin{align}}
\newcommand{\eal}{\end{align}}

\newcommand{\bw}{\begin{widetext}}
\newcommand{\ew}{\end{widetext}}

\begin{document}

\title{Deflection angle of charged massive particles in slowly rotating Kerr-Newman space-times via  Gauss-Bonnet theorem and Hamilton-Jacobi method }

\author{Kimet Jusufi}
\email{kimet.jusufi@unite.edu.mk}
\affiliation{Physics Department, State University of Tetovo, Ilinden Street nn, 1200,
Tetovo, Macedonia.}
\affiliation{Institute of Physics, Faculty of Natural Sciences and Mathematics, Ss. Cyril and Methodius University, Arhimedova 3, 1000 Skopje, Macedonia.}

\begin{abstract}
In this paper the problem of gravitational deflection of relativistic charged massive particles in slowly rotating Kerr-Newman black holes is considered. Toward this purpose we have used two methods: Firstly, we have applied the Gauss-Bonnet theorem (GBT) and the optical geometry to evaluate the deflection angle of charged particles. Secondly, we have presented a detailed analysis of the deflection angle by means of the Hamilton-Jacobi equation recovering the same invariant result for the deflection angle in leading order terms. The crucial point behind the first method is to use the correspondence between the motion of a charged massive particles in a gravitational field in presence of electromagnetic field and the motion of photons in a non-homogeneous cold plasma. It is shown that the deflection angle besides the black hole (BH) mass $M$, BH angular momentum $a$, BH electric charge $Q^2$, BH magnetic charge $P^2$, it is also affected by the electric charge of the particle $q$ and relativistic velocity of the particle $v$.  

\end{abstract}


\date{\today}
\maketitle

\section{Introduction}

Deflection of light by a massive body led to the first experimental proof of the theory of general  
relativity \cite{Eddington}. Usually the calculations of the deflection angle of light are based on the standard formulation of geodesic equations. In this direction one can study the weak field approximation and
strong gravitational lensing \cite{Schneider, Darwin,Virbhadra:2007kw,Virbhadra:2008ws,Eiroa:2002mk,Zhao:2015fya,Nandi:2006ds,Tsukamoto:2017hva,Shaikh:2017zfl,Gyulchev:2006zg,Aliev:2009cg,Tsukamoto:2017edq,iyer}. More recently, the phenomenon of gravitational lensing was important in the Event Horizon Telescope (EHT) Collaboration with the announcement of the first image of a supermassive black hole at the center of a M87 galaxy \cite{m87, Akiyama:2019eap}. Basically, if a black hole is surrounded by the accretion disk the image of such a black hole with appear distorted due to the gravitational lensing. In particular, by studying the light-like geodesics one can argue that the photons can be absorbed or can escape from the black hole resulting with a dark region known as the shadow. In this way, the black hole shadow has become a tool and a great opportunity in astrophysics to study the nature of black hole in the strong gravity regime in the future.

Recently, in a seminal paper, Gibbons and Werner showed that the deflectio angle of light in the weak limit can be viewed as a global effect \cite{Gibbons}. Namely, they considered the source and the observer to be located at theasymptotic region, then by appling the GBT to the optical metric of a lens the exact deflection angle is recovered for static metrics. Consequently this method was extend by Werner \cite{Werner} to study the deflection of light by a Kerr black hole using the Kerr-Randers optical geometry. 
This method was extended to compute the deflection of 
light by cosmic strings/global monopoles which are charachterized by a nontrivial spacetime topology, but also for rotating and non-rotating wormholes in Refs. \cite{Jusufi11,Jusufi22,Jusufi33,Jusufi44}.
An important contribution was made by Ishihara et al \cite{ishihara1,ishihara2} and Ono et al \cite{ishihara3,ishihara4} which, among other things, they obtained the bending angle of light by using the GBT by introducing a finite distance correction from a source and observer.

More recently the deflection of massive particles in the gravitational field attrected a lot of interest. Namely, Crisnejo et al \cite{Crisnejo1,Crisnejo2,Crisnejo3,Crisnejo4} showed that how one can successfully 
study the deflection of massive particles using the GBT theorem in a plasma medium for a static and spherically symmetric gravitational field such as the Schwarzschild and Reissner-Nordstrom spacetimes. Among other things, they pointed out the correspondence between the non-geodesic motion of test charged massive particles in a gravitational field and the motion of photons in a non-homogeneous cold non-magnetized plasma in the Reissner-Nordstrom spacetime.  In Ref. \cite{Jusufi1}, Jusufi used a new method to compute the deflection angle for massive particles based on the isotropic type metrics for a slowly rotating Kerr black hole and Teo wormhole and the refractive index of the corresponding optical media appling the GBT. In this method, particles can be considered as a de Broglie wave packets. Furthermore in Ref. \cite{Jusufi2}, it was shown that one can use the deflection angle of massive particles to distinguish different objects such as a rotating naked singularities from Kerr-like wormholes. Other contribution include deflection angle of neutral massive particles using standard methods, such as in Schwarzschild geometry \cite{Jia}, Schwarzschild black hole in radiation gauge \cite{Li:2019pvi}, in Reissner-Nordstrom spacetime \cite{Pang}, Kerr spacetime \cite{kerr1,kerr2,Hackmann:2013pva}. From a physical point of view, deflection of massive particles can be used for example to investigate the lensing of neutrinos and cosmic rays emitted by these supernova \cite{Yu:2003hj,Brown:2005ta,Patla}.

To our best knowledge, the problem of computing the deflection angle of charged massive particles in stacionary spacetimes such as the Kerr-Newman black has not been solved yet. In this paper, our aim is to extend the method reported by Crisnejo et al \cite{Crisnejo1} to calculate the deflection angle of charged massive particles in a Kerr-Newman spacetime.  More specifically we shall use the correspondence between the non-geodesic motion of a charged massive particles in a gravitational field with electric field and the motion of photons in a non homogeneous plasma pointed out in \cite{Crisnejo1}.

The paper is organized as follows. In Sec. I we use shall calculate the deflection angle of massive charged particles in a slowly rotating black hole. We shall apply the GBT to the optical geometry. In Sec. II, we consider the problem of the gravitational deflection of massive charged particles using the Hamilton-Jacobi approach. In Sec. III. we consider the effect of BH magnetic charge on the deflection angle. Finally, in Sec. IV, we comment on our results. We shall use a geometrized unit with $G=c=\hbar= 1$ throughout this paper.

\section{Deflection of massive charged particles in Kerr-Newman spacetimes}
Let us start by recalling the Kerr-Newman solution in the Boyer-Lindquist form given as follows
\begin{equation}
ds^2=\left(1-\frac{2M r-Q^2}{\Sigma^2}\right)dt^2-\frac{\Sigma^2}{\Delta}dr^2-\Sigma^2 d\theta^2-\sin^2\theta d\phi^2 
\left(r^2+a^2+\frac{(2Mr-Q^2)}{\Sigma^2}a^2\sin^2 \theta \right)+2a \sin^2 \theta \frac{2Mr-Q^2}{\Sigma^2}d\phi dt
\end{equation}
with 
\begin{equation}
\Sigma^2=r^2+a^2 \cos^2\theta,
\end{equation}
\begin{equation}
\Delta=r^2-2Mr+a^2+Q^2.
\end{equation}
Note that $M$ is the black hole mass, $Q$ is the electric charge, and $a$ is the angular momentum parameter. The metric (1) linearized  in $a$ can be written as 
\begin{equation}
ds^2=\left(1-\frac{2M}{r}+\frac{Q^2}{r^2} \right)dt^2+\frac{2a \sin^2 \theta \left(2Mr-Q^2\right)}{r^2} d\phi dt-r^2  d\theta^2-\frac{dr^2}{1-\frac{2M}{r}+\frac{Q^2}{r^2}}-r^2\sin^2 \theta d\phi^2.
\end{equation}

We shall be interested to study the deflection of charged particles in the spacetime element (4). Firstly, to simplify the problem, we can consider the deflection in the equatorial plane with $\theta=\pi/2$. Secondly, we can simplify the problem further if we rewrite the last metric in the following form
\begin{equation}
ds^2=A(x^i)dt^2-g_{ij}dx^i dx^j=A(x^i)dt^2-\frac{dr^2}{1-\frac{2M}{r}+\frac{Q^2}{r^2}}-r^2d\phi^2,
\end{equation}
where $i,j=(r,\phi)$. Moreover
\begin{equation}
A(r)=1-\frac{2M}{r}+\frac{Q^2}{r^2}+\frac{2a\left(2Mr-Q^2\right)}{r^2}\frac{d\phi}{dt}.
\end{equation}

In Ref. \cite{Crisnejo1} it has been argued that there exists a correspondence between the non-geodesic motion of test charged massive particles in a gravitational field with electrical field and the motion of photons in a non homogeneous cold non-magnetized plasma. The Hamiltonian of a charged particle in a curved spacetime (5) takes the form  \cite{Crisnejo1} 
\begin{equation}
    \mathcal{H}(x^i,p_i)=\sqrt{\mu^2A(x^i)+A(x^i)g^{ij}p_i p_j}+qU(x^i).
\end{equation}

More specifically, one can use the Jacobi metric formulation to study the motion of test massive particles in spacetime (5)  derived from a given Hamiltonian as shown by Gibbons \cite{Gibbons:2015qja}. In particular it is shown that the motion of charged massive particles is given by the geodesics of a energy-dependent Riemmanian metric known as the Jacobi metric  $J_{ij}$ given by  \cite{Crisnejo1},
\begin{equation}
    J_{ij}=E^2 \; \hat{g}_{ij}^{\text{opt}},
\end{equation}
where
\begin{equation}\label{analog-prefactor}
    \hat{g}_{ij}^{\text{opt}}= \bigg[ \bigg( 1-\frac{qU(x^i)}{E} \bigg)^2 -\frac{\mu^2 A(x^i)}{E^2} \bigg] \frac{g_{ij}}{A(x^i)}.
\end{equation}
with $U(x^i)=Q/r$. In the special case for the massless test particle, $\hat{g}_{ij}^{\text{opt}}$ coincides with the optical metric $g_{ij}^{\text{opt}}$. On the other hand, the photon trajectory in a medium described by a refractive index $n(r)$ are geodesics with respect to the optical metric \cite{Crisnejo1}
\begin{equation}\label{eq:optmcp}
g_{ij}^{\text{opt}}= n^2(x^i)\frac{g_{ij}}{A(x^i)},
\end{equation}

Let us now rewrite the corresponding optical metric to (5) which has the following form
\begin{equation}
d\sigma^2=g^{\text{opt}}_{ij} dx^i dx^j=n^2(r)\left(\frac{dr^2}{\left(1-\frac{2M}{r}+\frac{Q^2}{r^2}\right)\left(1-\frac{2M}{r}+\frac{Q^2}{r^2}+\frac{2a\left(2Mr-Q^2\right)}{r^2}\frac{d\phi}{dt} \right)}+\frac{r^2 d\phi^2}{\left(1-\frac{2M}{r}+\frac{Q^2}{r^2}+\frac{2a\left(2Mr-Q^2\right)}{r^2}\frac{d\phi}{dt} \right)}\right),
\end{equation}
then using the above correspondence, for the refractive index $n$ it follows
\begin{equation}
n^2(r)= \left(1-\frac{ q Q }{r E}\right)^2-\frac{\mu^2}{E^2}A(r).
\end{equation}

We can use a further simplified result using the relativistic velocity and angular momentum of the particle 
\begin{equation}
E=\frac{\mu}{\sqrt{1-v^2}}\ \ \ \ \text{and} \ \ \ J=\frac{\mu v b}{\sqrt{1-v^2}},
\end{equation}
where $\mu$ is the rest mass of the particle. It follows that in leading order of $q$
\begin{equation}
n^2(r)\simeq 1-\frac{2 q Q }{r E}-(1-v^2)A(r).
\end{equation}

Our next goal is to calculate the quantity $d\phi/dt$. To to so, we recall that the Lagrangian density  of a test charged particle in a curved spacetime (4) is given by
\begin{equation}
\mathcal{L}=\frac{1}{2}g_{\mu \nu}\dot{x}^{\mu}\dot{x}^{\nu}+qA_{\mu}\dot{x}^{\mu}
\end{equation}

From the symmetries one can obtain two constants of motion corresponding to conservation of energy, $E$, and angular momentum  $J$, respectively. Therefore, using metric (4) we have
\begin{equation}\label{ConservationEL}
\begin{split}
    p_{t}&=\left(1-\frac{2M}{r}+\frac{Q^2}{r^2} \right)\dot{t}+\frac{a\left(2Mr-Q^2\right)}{r^2}\dot{\phi}+qA_{t}=E, \\
    p_{\phi}&=-r^2\dot{\phi}+\frac{a \left(2Mr-Q^2\right)}{r^2}\dot{t}+q A_{\phi}=-J.
\end{split}
\end{equation}
where 
\begin{equation}
A_{t}=\frac{Q}{r},\,\,\,\,\,\,\,\, A_{\phi}=-\frac{a Q}{r}.
\end{equation}

Using Eq. (16) and $J/E=v b$, we  obtain
\begin{equation}
\begin{split}
 \frac{d\phi}{dt}=-\frac{\frac{a\left(2Mr-Q^2\right)}{r^2}+\left(1-\frac{2M}{r}+\frac{Q^2}{r^2} \right)\,v b-q\left[\frac{a\left(2Mr-Q^2\right)}{r^2}A_{t}-\left(1-\frac{2M}{r}+\frac{Q^2}{r^2} \right)A_{\phi}\right]}{-r^2+\frac{a\left(2Mr-Q^2\right)}{r^2}\,v b-q(-r^2\,A_{t}-\frac{a\left(2Mr-Q^2\right)}{r^2} A_{\phi})}.
\end{split}
\end{equation}

Using these results we can precede to to apply the GBT. Let us chose a non-singular region $\mathcal{D}_{R}$ with boundary $\partial
\mathcal{D}_{R}=\gamma _{\tilde{g}}\cup C_{R}$, then the GBT provides a relation between the geometry and the topology in terms of the following relation
\begin{equation}
\iint\limits_{\mathcal{D}_{R}}\mathcal{K}\,\mathrm{d}S+\oint\limits_{\partial \mathcal{%
D}_{R}}\kappa \,\mathrm{d}\sigma+\sum_{i}\theta _{i}=2\pi \chi (\mathcal{D}_{R}).
\end{equation}

Where $\kappa$ is known as the geodesic curvature, while $\mathcal{K}$ is the Gaussian optical curvature. 
In the case of a non-singular domain we know that the Euler characteristic number is one, i.e., 
$\chi (\mathcal{D}_{R})=1$. 
The geodesic curvature $\kappa$, can be defined as follows \cite{Gibbons}
\begin{equation}
\kappa =g^{op}\,\left(\nabla _{\dot{\gamma}}\dot{\gamma},\ddot{\gamma}\right),
\end{equation}
with the unit speed condition $g^{op}(\dot{\gamma},\dot{%
\gamma})=1$, in which $\ddot{\gamma}$ is the unit acceleration vector. By construction, there are two corresponding jump angles in the limit $R\rightarrow \infty $, resulting with $\theta _{\mathit{O}%
}+\theta _{\mathit{S}}\rightarrow \pi $. With these informations in hand, the GBT is simplified as follows
\begin{equation}
\lim_{R\rightarrow \infty }\int_{0}^{\pi+\hat{\alpha}}\left[\kappa \frac{\mathrm{d} \sigma}{\mathrm{d} \varphi }\right]_{C_R} \mathrm{d} \varphi =\pi-\lim_{R\rightarrow \infty }\iint\limits_{\mathcal{D}_{R}}\mathcal{K}\,\mathrm{d}S.
\end{equation}
Using the fact that there is a zero contribution from the geodesics i.e. $\kappa (\gamma_{\tilde{g}})=0$, we shall seek a contribution due to the curve $C_{R}$. This contribution can be calculated via 
\begin{equation}
\kappa (C_{R})=|\nabla _{\dot{C}_{R}}\dot{C}_{R}|~.
\end{equation}

Let us chose $C_{R}:=r(\varphi)=R=\text{const}$, where $R$ donates the distance from the coordinate origin. The radial part yields
\begin{equation}
\left( \nabla _{\dot{C}_{R}}\dot{C}_{R}\right) ^{r}=\dot{C}_{R}^{\phi
}\,\left( \partial _{\phi }\dot{C}_{R}^{r}\right) +\bar{\Gamma} _{\phi
\phi }^{r}\left( \dot{C}_{R}^{\phi }\right) ^{2}~. 
\end{equation}

Doing the correspondent identifications between frequencies and energy and
mass, it follows the form invariant quantity
\begin{equation}
\lim_{R\to \infty}\left[\kappa \frac{\mathrm{d} \sigma}{\mathrm{d} \phi }\right]_{C_R}=1
\end{equation}

Resulting in this way the form invariant deflection angle
\begin{equation}
\hat{\alpha}=-\int\limits_{0}^{\pi}\int\limits_{\frac{b}{\sin \varphi}}^{\infty}\mathcal{K} dS.
\end{equation}

This equation encodes the global effect on the lensing of particles due to the fact that one has to integrate over the optical domain of integration outside the enclosed mass. The Gaussian optical curvature for the two-dimensional optical metric (11) can be computed by making use the definition
\begin{equation}
\mathcal{K}=\frac{\mathcal{R}}{2}
\end{equation}
where $\mathcal{R}$ is the Ricci scalar for the optical metric. In particular for the Gaussian optical curvature of the optical metric (11) we find the following  expression
\begin{equation}
\mathcal{K}=\frac{2 A''n^2 A B r-4 n''n A^2 Br-2 n^2 B A''r+An^2 (r B'+2 B)A'+(-2 n(r n'+n) B'+B n'(r n'-n))A^2}{4 n^4 A r},
\end{equation}
note that $'$ donates derivation with respect to $r$ and
\begin{equation}
B(r)=1-\frac{2M}{r}+\frac{Q^2}{r^2}.
\end{equation}

The Gaussian optical curvature approximated in leading order terms yields
\begin{equation}
\mathcal{K}\simeq -\frac{M(1+v^2)}{r^3 v^4}+\frac{Q^2(2+v^2)}{v^6 r^4}+\frac{q Q }{r^3 E v^4}+\frac{18 b M a}{v^3 r^5}.
\end{equation}

Using the last result from (25) the deflection angle reads 
\begin{equation}
\hat{\alpha}=-\int\limits_{0}^{\pi}\int\limits_{\frac{b}{\sin \varphi}}^{\infty}\left(-\frac{M(1+v^2)}{r^3 v^4}+\frac{Q^2(2+v^2)}{v^6 r^4}+\frac{q Q }{r^3 E v^4}+\frac{18 b M a}{v^3 r^5}\right)dS.
\end{equation}
With the surface optical element
\begin{equation}
dS=\sqrt{\det g^{op}} dr d\phi=\left[\frac{v^2 r EA(r)-r E A(r)+r E-2q Q}{EA(r)\sqrt{B(r)}}\right]dr d\phi.
\end{equation}
In leading order terms it can be shown $
dS  \simeq r v^2 dr d\phi.
$. This integral can easily be evaluated, yielding
\begin{equation}
\hat{\alpha}\simeq \frac{2M}{b}\left(1+\frac{1}{v^2}\right)-\frac{\pi  Q^2}{4 b^2}\left(1+\frac{2}{v^2}\right)-\frac{2 qQ }{b v^2 E} \pm \frac{4aM }{b^2 v}.
\end{equation}
Let us mention here that the positive and negative sign is for a prograde and retrograde case, respectively. Namely, $a>0$ means that BH is co-rotating relative to the observer, while for $a<0$ it is counter-rotating relative to BH. 
Setting $q=0$, we recovered the gravitational deflection angle of neutral massive particles in a Kerr-Newman geometry.  Letting $a=0$, our result is in perfect agreement with the result reported in Ref. \cite{Crisnejo1}. Moreover in the special case, $v=c=1$ the light deflection angle is recovered.

\section{Geodesics Approach}
In this section, we will investigate the geodesic equation of the charged massive particle  with rest mass $\mu$ and electric charge $q$. Our aim is to calculate the deflection angle of the charged particle in a slowly rotating Kerr-Newman geometry.
In particular we will make use of the Hamilton-Jacobi equation which can be written as
\begin{equation}\label{HJE}
    -\frac{\partial S}{\partial \sigma}=\frac{1}{2}g^{\mu\nu}\left(\frac{\partial S}{\partial x^{\mu}}+q A_{\mu}\right)
    \left(\frac{\partial S}{\partial x^{\nu}}+q A_{\nu}\right),
\end{equation}
where $S$ is the Jacobi function and $\sigma$ a geodesic affine parameter. Hereafter, we will consider only a light ray laying on $\theta=\pi/2$ in the spacetime metric (4). In addition, we have the constant of motions corresponding to the 
conservation of the the energy $E$ and the angular momentum of the particle 
$J$ measured at infinity, respectively. Therefore, we shall seek a solution of the Hamilton-Jacobi equation of the form
\begin{equation}\label{SolutionForm}
    S=-\frac{1}{2}\mu^2 \sigma +E\,t -J\,\phi +S_{r}(r,E,J).
\end{equation}

Before continuing we note that we shall be interested only in terms which are linear in $q$. The Hamilton-Jacobi equation than takes the form
\begin{equation}\label{HJE2}
    \frac{1}{g_{rr}}\left(\frac{d S_{r}(r)}{d r}\right)^2+\frac{g_{\phi\phi}E^2+2g_{t\phi}EJ+g_{tt}J^2+2q(g_{\phi \phi}E A_t-g_{tt}A_{\phi}J)-2 q g_{t \phi}(E A_t-J A_{\phi})}{g_{tt}g_{\phi\phi}-g_{t\phi}^2}=\mu^2.
\end{equation}
with
\begin{eqnarray}
g_{tt} &= &1-\frac{2M}{r}+\frac{Q^2}{r^2}, \\
g_{t \phi}& = &\frac{a  \left(2Mr-Q^2\right)}{r^2}, \\
g_{rr} &= &-\left(1-\frac{2M}{r}+\frac{Q^2}{r^2}\right)^{-1}, \\
g_{\phi \phi} & = & -r^2.
\end{eqnarray}
Using the relation $p_{r}=g_{rr}\dot{r}=\partial S/\partial r=dS_{r}/dr$, the radial equation of motion is obtained as follows
\begin{equation}\label{RadialEqMotion}
\begin{split}
    \left(\frac{dr}{d\sigma}\right)^2 & =\frac{g_{\phi\phi}E^2+2g_{t\phi}EJ+g_{tt}J^2+2q(g_{\phi \phi}E A_t-g_{tt}A_{\phi}J)-2 q g_{t \phi}(E A_t-J A_{\phi})}{g_{rr}(g_{t\phi}^2-g_{tt}g_{\phi\phi})}+\frac{\mu^2}{g_{rr}} \\
              & =\frac{g_{\phi\phi}(E-V_{+})(E-V_{-})}{g_{rr}(g_{t\phi}^2-g_{tt}g_{\phi\phi})}+\frac{\mu^2}{g_{rr}},
\end{split}
\end{equation}
where the solutions of the equation in $E$
\begin{equation}\label{EffPotentialEq}
   g_{\phi\phi}E^2+2g_{t\phi}EJ+g_{tt}J^2+2q(g_{\phi \phi}E A_t-g_{tt}A_{\phi}J)-2 q g_{t \phi}(E A_t-J A_{\phi})=0,
\end{equation}
represent in terms of the effective potentials
\begin{equation}\label{EffPotensials}
    V_{\pm}(r)=\frac{1}{g_{\phi\phi}}\bigg(qA_t(g_{t\phi}-g_{\phi \phi})-Jg_{t\phi}\pm\sqrt{J^2(g_{t\phi}^2-g_{tt}g_{\phi\phi})-2 q J (A_t g_{t \phi}^2-g_{t \phi} g_{\phi \phi} (A_t-A_{\phi})-g_{tt} g_{\phi \phi}A_{\phi})}\bigg).
\end{equation}

At this point one can evaluate the Hamiltonian at the closest distance where $\dot{r} = 0$, from where we find the precise relation between the impact parameter $\xi=J/E=wb$ and the closest approach distance $r_{0}$
\begin{equation}\label{ImpactParameterEq}
    \xi(r_0)=\frac{ \frac{q}{E}A_{\phi}(g_{tt}-g_{t \phi})-g_{t\phi}+\sqrt{(g_{t\phi}^2-g_{tt}g_{\phi\phi})[1-g_{tt}(1-v^2)]}}{g_{tt}}\Big{|}_{r_{0}}.
\end{equation}

This result reduces to photon impact parameter $b$ when $q=0$ and $v=c=1$. The sign in front is related to the fact that the the particle can co-rotate or counter-rotate relative to the black hole. Next, the orbital equation of the particle in terms of azimuthal $\phi$ and radial $r$ coordinates can be obtained by using the fact that the partial derivative of the Jacobi function with respect to the constant of motion $J$ is a constant
\begin{equation}\label{EqMotionPhir}
    \frac{\partial S}{\partial J}=\phi-\frac{\partial S_{r}(r,E,J)}{\partial J}=\text{const}.
\end{equation}

Solving Eq. (\ref{HJE2}) and taking into account Eq. (\ref{RadialEqMotion}), we obtain the azimuthal shift of the charged particle as follows
\begin{equation}\label{OrbitalEq}
\begin{split}
    \phi(r)-\phi(r_{0}) & =\frac{\partial S_{r}(r,E,J)}{\partial J} \\
        & =\quad \int_{r_{0}}^{r}g_{rr}\frac{\partial}{\partial J}\left(\frac{dr}{d\sigma}\right)dr \\
        & =\pm\int_{r_{0}}^{r}\frac{(g_{t\phi}E+g_{tt}J+2 q A_{\phi}(g_{t \phi}-g_{tt}))}{(g_{t\phi}^2-g_{tt}g_{\phi\phi})}  \left[\frac{g_{\phi\phi}(E-V_{+})(E-V_{-})}{g_{rr}(g_{t\phi}^2-g_{tt}g_{\phi\phi})}+\frac{\mu^2}{g_{rr}}\right]^{-1/2}dr \\
            & =\pm\int_{r_{0}}^{r}\frac{(g_{t\phi}+\xi g_{tt}+ \frac{2q}{E} A_{\phi}(g_{t \phi}-g_{tt}))}{(g_{t\phi}^2-g_{tt}g_{\phi\phi})}  \left[\frac{g_{\phi\phi}(1-\mathcal{V}_{+})(1-\mathcal{V}_{-})}{g_{rr}(g_{t\phi}^2-g_{tt}g_{\phi\phi})}+\frac{1-v^2}{g_{rr}}\right]^{-1/2}dr.
\end{split}
\end{equation}

Note that $\mathcal{V}_{\pm}(r,\xi)\equiv V_{\pm}(r,J)/E$ gives the  effective potentials per unit of the energy and it is related to the impact parameter $b$. With these results in hand, we can calculate the deflection angle of the charged particle by assuming that the source $r_{S}$ and observer $r_{O}$ are placed in the asymptotically flat region of the spacetime, in other words $r_{S}\rightarrow\infty$ and $r_{O}\rightarrow\infty$. Taking $\xi=v\,b$ and changing to new variable $r=1/u$, we obtain
\begin{equation}\label{OrbitalEq}
\begin{split}
   \phi_{u=0}-\phi_{u=1/b}&=\int_{0}^{1/b}\frac{(g_{t\phi}+v b g_{tt}+ \frac{2q}{E} A_{\phi}(g_{t \phi}-g_{tt}))}{u^2(g_{t\phi}^2-g_{tt}g_{\phi\phi})}  \left[\frac{g_{\phi\phi}(1-\mathcal{V}_{+})(1-\mathcal{V}_{-})}{g_{rr}(g_{t\phi}^2-g_{tt}g_{\phi\phi})}+\frac{1-v^2}{g_{rr}}\right]^{-1/2}du,
\end{split}
\end{equation}

The final deflection angle of massive particles can be summarized as follows
\begin{eqnarray}\label{DA}
    \hat{\alpha}&=& 2|\phi_{u=0}-\phi_{u=1/b}|-\pi.
\end{eqnarray}

If we consider a series expansion in (46) and then evaluating the integration we recover the following relation for the deflection angle 
\begin{equation}
\hat{\alpha}\simeq \frac{2M}{b}\left(1+\frac{1}{v^2}\right)-\frac{\pi  Q^2}{4 b^2}\left(1+\frac{2}{v^2}\right)+\frac{2 qQ }{b v^2 E} +\frac{3 \pi q M Q}{b^2 E \,v^2}\pm \frac{4aM }{b^2 v}\pm \frac{Q^2 a \pi}{b^3 v}.
\end{equation}

Finally making the identification $q \to -q$, one finds
\begin{equation}
\hat{\alpha}\simeq \frac{2M}{b}\left(1+\frac{1}{v^2}\right)-\frac{\pi  Q^2}{4 b^2}\left(1+\frac{2}{v^2}\right)-\frac{2 qQ }{b v^2 E} -\frac{3 \pi q M Q}{b^2 E \,v^2}\pm \frac{4aM }{b^2 v}\pm \frac{Q^2 a \pi}{b^3 v}.
\end{equation}

The last result in leading order terms is in perfect agreement with Eq. (32). In general, depending on the sign of electric charge, i.e. $q \to \pm q$ we have two results; namely Eq. (48) and Eq. (49). Physically this term describes the interaction between the charged black hole and the charged particle. Note that terms like $3 \pi q M Q/b^2 E \,v^2$ and  $\pm Q^2 a \pi/b^3 v$ can be viewed as a second order terms. In order to find these terms in the GBT one has to modify the the equation for the particle $r(\phi)$ in the integration domain. We leave this problem as a future project.

\section{Effect of magnetic charge on the deflection angle  by Kerr-Newman black holes}
One can extend these results by considering also the effect of magnetic charge. In particular if we include also the magnetic charge $P$ of the gravitating source in the Kerr-Newman black hole then the solution is given by
\begin{eqnarray}\notag
ds^2&=&\left(1-\frac{2M r-(Q^2+P^2)}{\Sigma^2}\right)dt^2-\frac{\Sigma^2}{\Delta}dr^2-\Sigma^2 d\theta^2-\sin^2\theta d\phi^2 
\left(r^2+a^2+\frac{(2Mr-(Q^2+P^2))}{\Sigma^2}a^2\sin^2 \theta \right)\\
&+& 2a \sin^2 \theta \frac{2Mr-(Q^2+P^2)}{\Sigma^2}d\phi dt,
\end{eqnarray}
with 
\begin{equation}
\Sigma^2=r^2+a^2 \cos^2\theta,
\end{equation}
\begin{equation}
\Delta=r^2-2Mr+a^2+Q^2+P^2.
\end{equation}

The electromagnetic potential is given by
\begin{equation}
A=A_{\mu} dx^{\mu}=\frac{Qr}{\Sigma^2}(dt-a \sin^2\theta d\phi)+\frac{P}{\Sigma^2}\cos\theta (a dt-(r^2+a^2)d\phi).
\end{equation}

Considering the equatorial plane $\theta=\pi/2$, this metric has the following form linearized  in $a$ 
\begin{equation}
ds^2=\left(1-\frac{2M}{r}+\frac{Q^2+P^2}{r^2} \right)dt^2+\frac{2a  \left(2Mr-(Q^2+P^2)\right)}{r^2} d\phi dt-\frac{dr^2}{1-\frac{2M}{r}+\frac{Q^2+P^2}{r^2}}-r^2d\phi^2.
\end{equation}

Introducing $\tilde{Q}^2_{m} \to Q^2+P^2$ in the last metric, it is straightforward to show that for the deflection angle it is obtained
\begin{equation}
\hat{\alpha}\simeq \frac{2M}{b}\left(1+\frac{1}{v^2}\right)-\frac{\pi  \tilde{Q}^2_{m}}{4 b^2}\left(1+\frac{2}{v^2}\right)-\frac{2 qQ }{b v^2 E} -\frac{3 \pi q M Q}{b^2 E \,v^2}\pm \frac{4aM }{b^2 v}\pm \frac{\tilde{Q}^2_{m} a \pi}{b^3 v}.
\end{equation}

We conclude that the magnetic charge $P^2$ also affects the deflection angle. In Fig. 1 we have plotted the deflection angle as a function of the impact parameter and black hole spin, respectively. It is shown that the bending angle increases with the decrease of the velocity for massive particles eventually approaching the deflection angle of light in the special case $v=c$. The right panel shows a linear increase of the bending angle relative to the black hole spin in the case of prograde case. 

\begin{figure*}
\includegraphics[width=8.4cm]{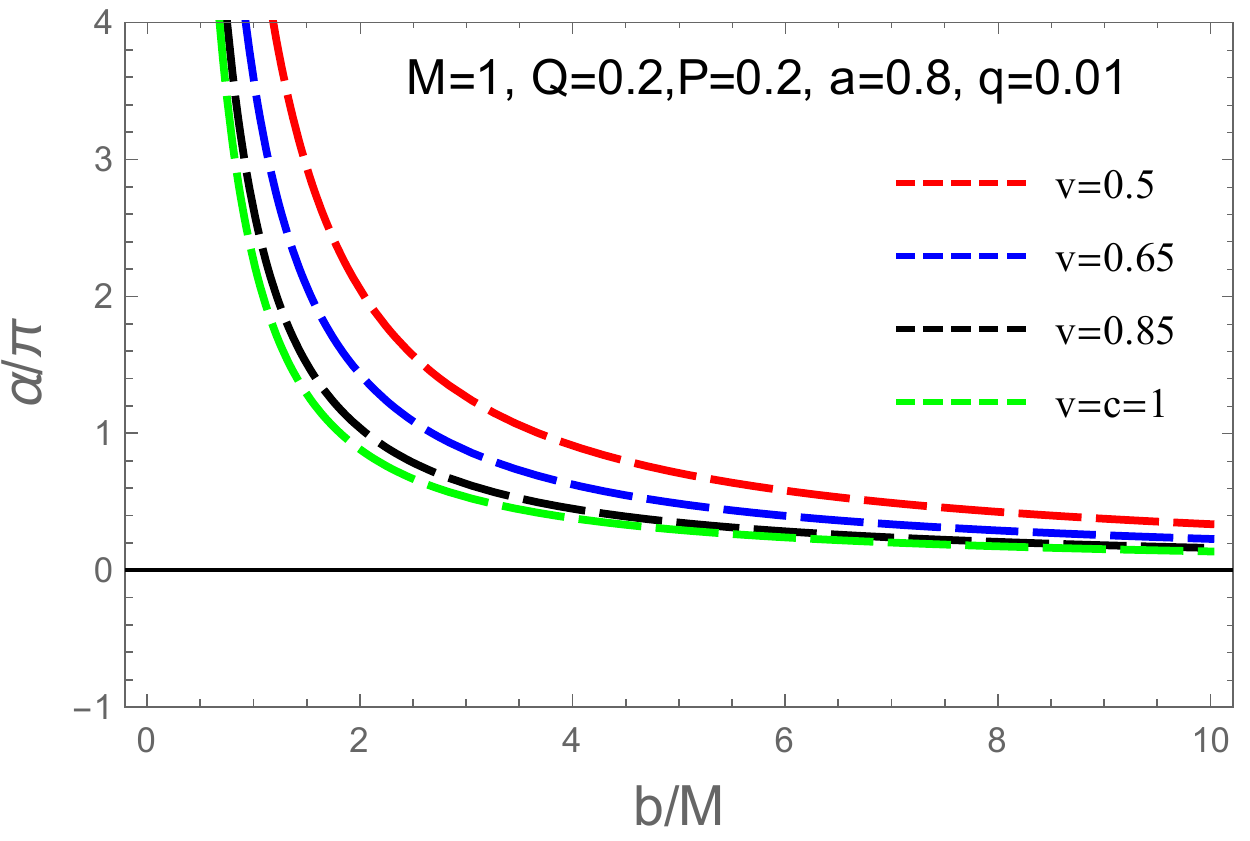}
\includegraphics[width=8.4cm]{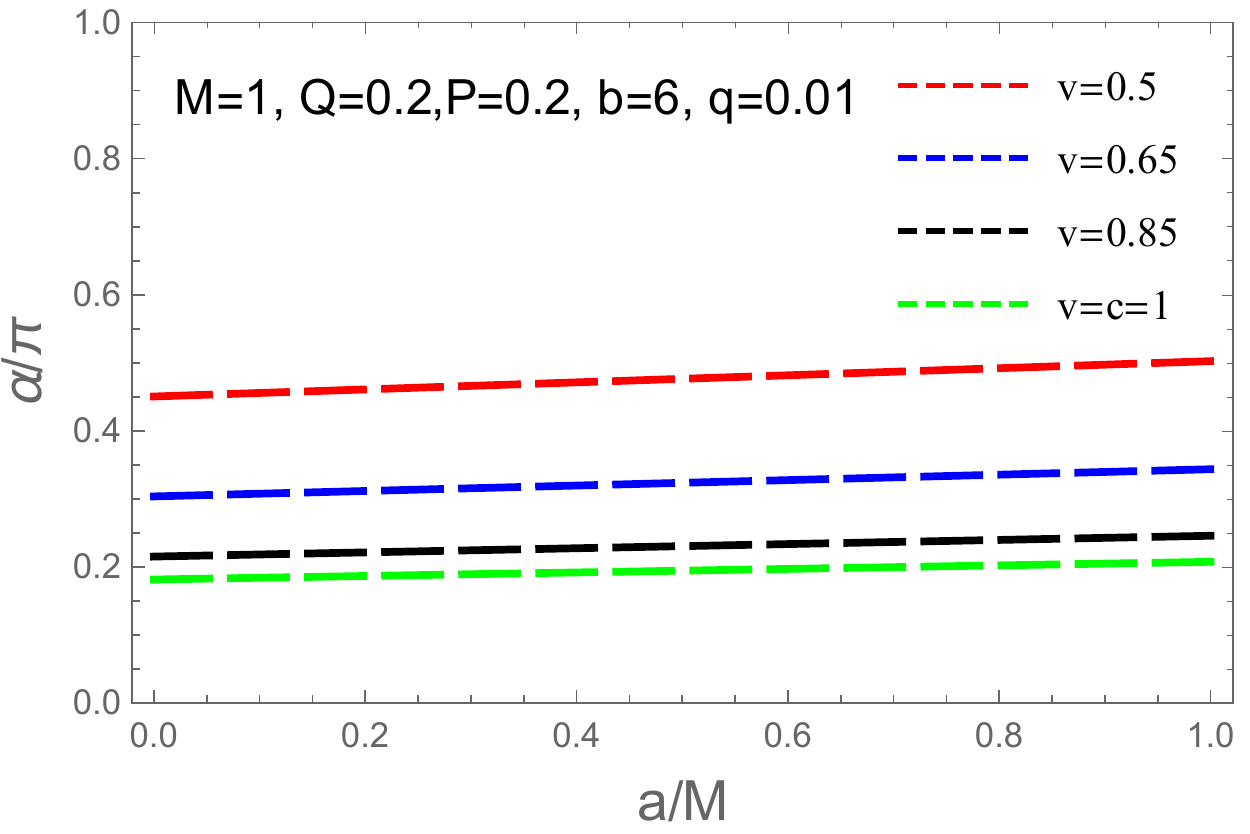}
\caption{Left panel: The deflection angle as a function of the impact parameter. Right panel: The deflection angle as a function of the angular momentum parameter. In both cases we observe that the deflection angle increases with a decrease of the relativistic velocity $v$. }
\end{figure*}
\newpage
\section{Conclusions}
In this paper, for the first time we studied the gravitational deflection of charged massive particles by a rotating Kerr-Newman spacetimes. In this way, we have extended resent results reported by Crisnejo et al \cite{Crisnejo1} to a rotating spacetimes. In doing so, we have  used the correspondence between the motion of a charged massive particles in a gravitational field with electric field and the motion of photons in a non homogeneous plasma. The deflection angle is then evaluated by applying the GBT theorem to the optical geometry.  Although the geodesic curvature $\kappa(C_R)$ and the optical metric are 
modified due the velocity of the particle, the form invariant quantity $\kappa(C_R) d\sigma/d\phi \to 1$ is recovered. In other words, this suggest that the expression for the deflection angle remains also form invariant, thus in order to compute the deflection angle of charged particles basically we need to integrate the Gaussian optical curvature over the optical geometry, hence the global effects are important. The total deflection angle is found to depend on the BH mass $M$, BH angular momentum $a$, BH electric charge $Q^2$ and BH magnetic charge $P^2$, it is also affected by the electric charge of the particle $q$ and relativistic velocity of the particle $v$. Furthermore we verified our result by presenting a detailed analysis in terms of the Hamilton-Jacobi method. It is shown that the bending angle increases with the decrease of the velocity for massive particles, and eventually approaching the deflection angle of light in the special case $v=c$. As a final remark, we note that the deflection angle is not valid for slowly moving particles, i.e. $v\to 0$, due to the apparent singularity in the deflection angle. In other words, both methods works perfectly well only for relativistic charged particles.

\end{document}